\documentclass[%
preprint,
showpacs,
amsmath,amssymb,
aps,prl,
]{revtex4-1}
\usepackage{booktabs}
\usepackage{tabularx}
\usepackage{braket}
\usepackage{float}
\usepackage{graphicx}% Include figure files
\usepackage{dcolumn}% Align table columns on decimal point
\usepackage{bm}% bold math
\usepackage{subfigure}
\usepackage{color}

\usepackage[colorlinks,citecolor = blue, linkcolor=red,hyperindex,CJKbookmarks]{hyperref}

\def\be{\begin{equation}}
\def\ee{\end{equation}}
\newcommand{\ba}{\begin{eqnarray}}
\newcommand{\ea}{\end{eqnarray}}

\definecolor{pink1}{RGB}{255,192,203}
\definecolor{pink2}{RGB}{255,130,180}
\definecolor{blue2}{RGB}{135,206,235}
\definecolor{blue1}{RGB}{176,224,230}

\def\x{{\mbox{\boldmath$x$}}}
\def\u{{\mbox{\boldmath$u$}}}

\def\g{{\mbox{\boldmath$g$}}}

\def\eps{{\epsilon}}

\newcommand{\bea}{\begin{eqnarray}}
\newcommand{\eea}{\end{eqnarray}}

\begin{document}
%\preprint{}

%\title{Unifying theory for turbulent internally heated convection (and turbulent horizontal convection)} 
\title{Scaling in internally heated convection: a unifying theory} 
\author{Qi Wang$^{1,2}$}
\author{Detlef Lohse$^{1,3}$}\email{d.lohse@utwente.nl}
\author{Olga Shishkina$^{3}$}\email{Olga.Shishkina@ds.mpg.de}

\affiliation{$^1$Physics of Fluids Group and Max Planck Center for Complex Fluid Dynamics, MESA+ Institute and J. M. Burgers Centre for Fluid Dynamics, University of Twente, P.O. Box 217, 7500AE Enschede, The Netherlands\\
$^2$Department of Modern Mechanics, University of Science and Technology of China, Hefei 230027, China\\
$^3$Max Planck Institute for Dynamics and Self-Organization, 37077 G\"ottingen, Germany}
%\thanks{}

\date{\today}

\begin{abstract}
We offer a unifying theory for turbulent purely internally heated convection, generalizing the unifying theories of Grossmann and Lohse (2000, 2001) for Rayleigh--B\'enard turbulence and of Shishkina, Grossmann and Lohse (2016) for turbulent horizontal convection, which are both based  on the 
splitting of the kinetic and thermal dissipation rates in respective boundary and bulk contributions.  
We obtain  the mean temperature of the system and the Reynolds number (which are the response parameters)
as function of the control parameters, namely the internal thermal driving strength (called, when nondimensionalized, the 
Rayleigh--Roberts number) and the Prandtl number. 
The results of the theory are consistent with our direct numerical simulations.
\end{abstract}

\maketitle

Thermally driven turbulence  is omnipresent in nature and technology. 
The thermal driving can be thanks to 
the temperature boundary conditions such as in Rayleigh--B\'enard convection -- 
a flow in a container heated from below and cooled from above \citep{ahl09,loh10,chi12} --  
or in horizontal convection \citep{hug08,shi16, ShiW2016} or vertical convection \citep{Shishkina2016c, ng2015, Ng2017, Ng2018}, where parts of the top, bottom, or sidewalls of the container are set at different temperatures. 
 However, the thermal driving  can also be thanks to internal heating, where the 
temperature field is driven by some forcing in the bulk. In many cases in nature, both ways of driving play a role 
at the same time. 
E.g., this holds for the Earth's mantle due to the driving through the hot inner core of the Earth and an 
additional driving due to the decay of radioactive material inside the core, producing heat \citep{houseman1988,bercovici1989,tackley1993,bunge1996,lay2008,moore2013,mallard2016,schubert2001}. 
Thus, in the Earth's mantle, about 10--20\% of the heat is transferred from the core, while the rest occurs due to the internal heating \citep{schubert2001}.
The internal heating dominates also in the atmosphere of Venus \citep{tritton1967convection,tritton1975}, which is heated up due to the absorption of sun light. 
One more example is the formation of Pluto's polygonal terrain, which is caused not only by convection of Rayleigh--B\'enard type \citep{mckinnon2016,trowbridge2016}, but also by internally heated convection \citep{vilella2017}.
And, of course, internally heated convection is relevant in many engineering applications, e.g., liquid-metal batteries \citep{kim2013liquid,xiang2017}.

To obtain a theoretical understanding of thermally driven turbulence including the cases of mixed thermal driving, it is mandatory to first understand the pure and well defined cases, namely on the one hand turbulent Rayleigh--B\'enard convection (RBC, exclusively driven by the heated and cooled plates), and on the other hand turbulent purely internally heated convection (IHC, exclusively driven by thermal forcing of the temperature field in the bulk), see \cite{roberts1967,goluskin2012,goluskin2016penetrative,vilella2018}. 
For the former case, Grossmann and Lohse (GL) have developed a unifying theory \citep{gro00,gro01,gro02,gro04,ste13},
with which the heat transfer and the degree of turbulence can quantitatively be described as function of the control parameters, in excellent agreement with the experimental and numerical data over a range of more than 7 orders of magnitude in the control parameters $Ra$ and $Pr$. 
Later this theory was also extended to horizontal convection \citep{shi16} (HC) and double diffusive convection \citep{yang2018}. 
GL arguments were also applied to IHC, to estimate the bulk temperature for small and moderate $Pr$ \citep{goluskin2012}. 
A complete theory, however, does not yet exist for purely internally heated convection.

The objective of the present work is to apply the reasoning of GL's theory to the case of purely internally heated convection and to develop 
a unifying theory for this case. In addition, we perform direct numerical simulations (DNS) of turbulent purely internally heated 
convection over a large range of control parameters and compare the DNS results with the theoretical predictions. 
The DNS are conducted in two dimensions (2D), as (i) the theory is based on Prandtl's equations,
  which are also 2D in spirit, as (ii) 2D and 3D thermally driven turbulence show very close analogies with respect to the integral quantities, in particular for large Prandtl numbers $Pr \ge 1$  \citep{poe13}, and as (iii) otherwise, due to unavoidable limitations in available CPU time, we could explore only a much smaller portion of the parameter space. 
  
In RBC, next to the geometric aspect ratio $\Gamma$ of the sample (the ratio between lateral and vertical extensions), the control parameters of the system are the temperature difference between top and bottom wall (in dimensionless form, the Rayleigh number) and the ratio between kinematic viscosity $\nu$ and thermal diffusivity $\kappa$, namely the Prandtl number $Pr = \nu/\kappa$. 
The response of the system consists of the heat flux from bottom to top (in dimensionless form, the Nusselt number $Nu$) and the degree of turbulence (in dimensionless form, the Reynolds number $Re$). 
In IHC, instead of the Rayleigh number, the dimensionless driving strength $Rr$ of the temperature field takes the role of the second control parameter, next to  $Pr$. 
It is often called Rayleigh--Roberts number (and that is why we use the abbreviation $Rr$) and will be defined below. 
The main response parameter, next to $Re$,
%%%the Reynolds number, 
is the mean temperature which the bulk achieves thanks to the internal driving. This is related to the heat fluxes into the top and bottom plates; note that they are different from each other.  
So the objective of this paper is to explain how the mean temperature and the Reynolds number in turbulent IHC depend on $Rr$ and $Pr$, for large enough aspect ratio $\Gamma$ of the sample.

The flow in IHC is confined between two parallel plates with distance $L$, with the gravitational acceleration $\g\equiv -g{\bf e}_z$ acting orthogonally to these plates. 
The underlying dynamical equations within the Boussinesq approximation are the compressibility condition $\partial_i u_i=0$, and
\begin{eqnarray}
\partial_t u_i  + u_j \partial_j u_i &=& -\partial_i p
+\nu \partial_j^2 u_i +\beta g
\delta_{iz} \theta,
\label{eq1}
\\
\partial_t \theta  + u_j \partial_j
  \theta &=&  \kappa \partial_j^2 \theta + \Omega , \label{eq2}
\end{eqnarray}
for the velocity field $\u (\x , t ) $, the kinematic pressure field $p(\x , t)$, and the reduced temperature field $\theta (\x , t)\equiv T(\x , t)-T_{plate}$. 
Here $T_{plate}$ is the temperature of both top and bottom plates, $\beta$ is the thermal expansion coefficient, $\delta_{ij}$ the Kronecker delta and $\Omega$ the constant bulk driving of the temperature field, which in non-dimensional form is called Rayleigh--Roberts number 
 \be
%  Rr = { \beta g L^5 \Omega \over \kappa^2 \nu  }. 
  Rr = { \beta g L^5 \Omega / (\kappa^2 \nu)  }. 
  \label{eq-r}
 \ee
The  equations (\ref{eq1}) and (\ref{eq2}) are supplemented by the boundary conditions (BCs) $u_i = 0$ and $\theta = 0$ at both plates.
Periodic BCs are used in the horizontal direction.
 
The main responses of the system can be expressed in terms of the mean temperature $\Delta \equiv \langle \theta(\x,t ) \rangle _V$
achieved in the system, where the average $\langle \cdot\rangle _V$ is over volume and time. The nondimensional form of this response parameter is
 \be
 \tilde \Delta \equiv \kappa \Delta / (\Omega L^2). 
 \label{tD}
 \ee
The other main nondimensional response parameter is the Reynolds number $Re = U L/\nu$, 
 with $U\equiv\sqrt{\left<\boldsymbol{u}^2\right>_{V}}$.
 
Obviously, due to the internal heating, the heat flux 
\be Q (z) \equiv \langle u_z \theta \rangle - \kappa \langle \partial_z \theta \rangle 
\label{eq-q}
\ee
(or in dimensionless form $\tilde Q (z) \equiv Q (z) / (\Omega L)$)  in the system is not  constant as in RBC, but depends on the height 
$z$. 
Here, $\langle \cdot\rangle$ means average in time and in a plane of constant $z$.
However, a simple integration of eq.\ (\ref{eq2}) over the full volume yields that 
the quantity 
\be
\tilde Q_0\equiv {z/L} - \tilde Q (z)
\label{eq-Q0}
\ee
is constant for all $z$ and equals
\be
\tilde Q_0  = -\tilde Q(z=0)= {\kappa \over \Omega L} \langle \partial_z \theta \rangle|_{z=0}\geq0.
\label{eq-n}
\ee 
$\tilde Q_0$ is thus a further dimensionless response parameter of the system. 
Eq. (\ref{eq-n}) implies that the dimensionless heat flux $\tilde Q$ is non-positive at $z=0$.
Applying eq. (\ref{eq-Q0}) at $z=L$ gives the dimensionless 
flux at $z=L$, 
\be
\tilde Q(z=L) = - {\kappa \over \Omega L} \langle \partial_z \theta \rangle|_{z=L}  = 1- \tilde Q_0 \ge0,
\label{eq-n2}
\ee
Relations (\ref{eq-n}) and (\ref{eq-n2}) immediately show that $0 \le \tilde  Q_0 \le 1$. 

As it is well known \citep{shr90,ahl09}, in RBC exact relations between the time and volume averaged thermal and kinetic dissipation rates,
$\epsilon_\theta \equiv \kappa \langle (\partial_i \theta (\x , t))^2\rangle_{V}$ and 
$\epsilon_u  \equiv  \nu \langle (\partial_i u_j (\x , t))^2\rangle_{V} $,
and the dimensionless control and response parameters $Nu$, $Ra$, and $Pr$ 
can be obtained from multiplying the thermal advection equation with $\theta (\x,t) $ and the 
Navier--Stokes equation with $u_i(\x, t)$ and subsequent Gauss integration and time and space averaging. 
Here we apply the same procedure to Eqs.~(\ref{eq2}) and (\ref{eq1}) and obtain
\begin{eqnarray}
\epsilon_\theta &=& \Omega \Delta  =  { L^2 \over \kappa} \Omega^2 \tilde \Delta = {\kappa \Delta^2 \over L^2}  \tilde \Delta^{-1}  
\label{ex-th},\\
\epsilon_u &=& {\nu^3 \over L^4} Rr Pr^{-2} \left( {1\over 2} - \tilde Q_0
\right). 
\label{ex-u}
\end{eqnarray}
As $\epsilon_u \ge 0 $ is non-negative by definition, we can now further restrain the magnitude of the dimensionless
heat flux through the bottom plate: $0 \le \tilde  Q_0 \le 1/2$. 
Just as the corresponding relations in RBC, also here, Eqs.~(\ref{ex-th}) and (\ref{ex-u}) 
relate the averaged thermal and kinetic  dissipation rates with the dimensionless control ($Rr$, $Pr$) 
and response ($\tilde \Delta$, $\tilde Q_0$) parameters.

\begin{figure}
\unitlength1truecm
\includegraphics[width=0.6\textwidth]{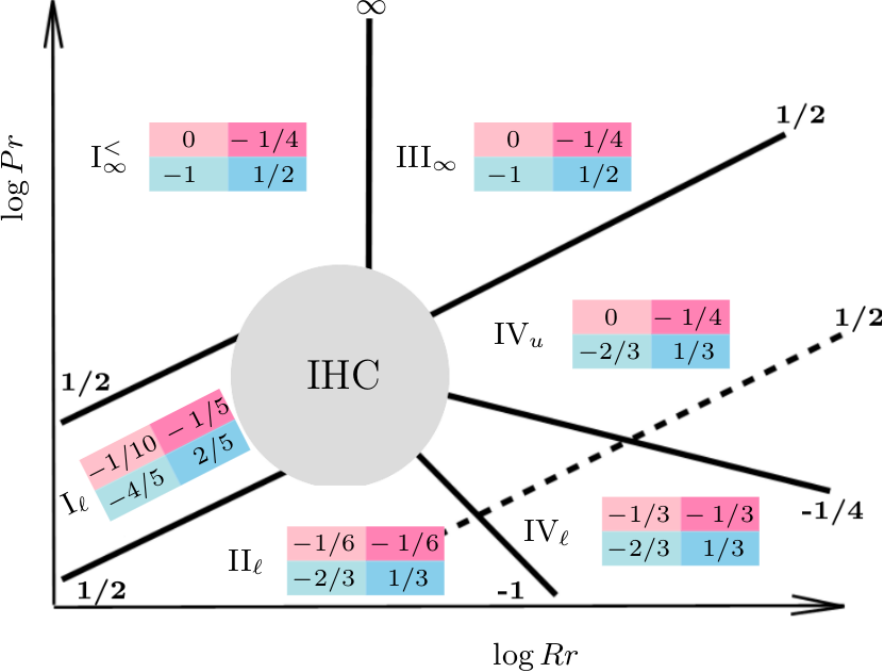}
\caption
{Sketch of the phase space of IHC with the various  limiting scaling regimes following from our unifying theory.
Continuous lines show the slopes of the transitions between the different regimes. The exponent $\beta_0$ of each slope $Pr\sim Rr^{\beta_0}$ between the different regimes in phase space 
is written in bold, close to the corresponding line.
The dashed  line shows the slope of the transition to the ultimate regime. 
For each regime, the color box represents the scaling exponents in the scalings of $\tilde\Delta$ and $Re$ versus $Pr$ and $Rr$:
$\left\{
 \begin{tabular}{l}\;\,$\tilde\Delta\sim Pr^{\footnotesize\fcolorbox{pink1}{pink1}{$\beta_1$}}Rr^{\footnotesize\fcolorbox{pink2}{pink2}{$\beta_2$}}$\\
 $Re\sim Pr^{\footnotesize\fcolorbox{blue1}{blue1}{$\beta_3$}}Rr^{\footnotesize\fcolorbox{blue2}{blue2}{$\beta_4$}}$
 \end{tabular}.\quad
\right.$ This  phase space for IHC is the analogous one to the one of  RBC of \cite{gro01} and the one of HC in \cite{shi16}. 
}
\label{Fig1}
\end{figure}

\begin{table}
\begin{tabular}{|c||c|c||c|c|}
\hline
Regime & \;$\epsilon_u / ( \nu^3 L^{-4} ) \sim Pr^{-2}Rr$\;&\;$\epsilon_\theta/(\kappa \Delta^2 L^{-2})\sim \tilde \Delta^{-1}$\;&\;$\tilde \Delta$\;&\;$Re$\;\\
\hline
I$_\infty^<$ &$\sim Re ^{2}$  &$\sim Pr^{1/2} Re^{1/2}$       &$\sim Pr^{0}\,Rr^{-1/4}$&$\sim Pr^{-1}\,Rr^{1/2}$\\
I$_u$        &$\sim Re ^{5/2}$&$\sim Pr^{1/3} Re^{1/2}$       &$\sim Pr^{1/15}\,Rr^{-1/5}$&$\sim Pr^{-4/5}\,Rr^{2/5}$\\
I$_\ell$     &$\sim Re ^{5/2}$&$\sim Pr^{1/2} Re^{1/2}$       &$\sim Pr^{-1/10}\,Rr^{-1/5}$&$\sim Pr^{-4/5}\,Rr^{2/5}$\\
II$_\ell$    &$\sim Re ^{3}$  &$\sim Pr^{1/2} Re^{1/2}$       &$\sim Pr^{-1/6}\,Rr^{-1/6}$&$\sim Pr^{-2/3}\,Rr^{1/3}$\\
III$_\infty$ &$\sim Re ^{2}$  &$\sim Pr\,Re\,\tilde\Delta$    &$\sim Pr^{0}\,Rr^{-1/4}$&$\sim Pr^{-1}\,Rr^{1/2}$\\
IV$_u$       &$\sim Re ^{3}$  &$\sim Pr\,Re^{3/2}\tilde\Delta$&$\sim Pr^{0}\,Rr^{-1/4}$&$\sim Pr^{-2/3}\,Rr^{1/3}$\\
IV$_\ell$    &$\sim Re ^{3}$  &$\sim Pr\,Re$                  &$\sim Pr^{-1/3}\,Rr^{-1/3}$&$\sim Pr^{-2/3}\,Rr^{1/3}$\\
\hline
\end{tabular}
\caption{Scalings of $\epsilon_u$, $\epsilon_\theta$, $Re$ and $\tilde \Delta$ in the different limiting regimes in IHC.}
\label{Tab1}
\end{table}

The key idea of the GL theory \citep{gro00,gro01} is to split the 
kinetic and thermal dissipation rates into contributions 
from the corresponding boundary layers (BLs) and bulks,
\begin{eqnarray*}
\epsilon_u=\epsilon_{u,\text{BL}}+\epsilon_{u, \text{bulk}},\quad
\epsilon_\theta={\,\epsilon_{\theta, \text{BL}}}+{\epsilon_{\theta, \text{bulk}},}
\end{eqnarray*}
and to apply the respective scaling relations for those 
(i.e., for 
$\epsilon_{u,\text{BL}}$,
$\epsilon_{u,\text{bulk}}$, 
$\epsilon_{\theta,\text{BL}}$ and
$\epsilon_{\theta,\text{bulk}}$), based on boundary layer theory and Kolmogorov's theory for fully developed turbulence in the bulk. 
The introduced scaling regimes I, II, III, and IV correspond to BL--BL, bulk--BL, BL--bulk, and bulk--bulk dominance in $\epsilon_u$ and $\epsilon_\theta$, respectively.
Here one should also take into account mean thicknesses of the thermal BLs ($\lambda_\theta$) and viscous BLs ($\lambda_u$).
The cases $\lambda_\theta<\lambda_u$ (large $Pr$) and $\lambda_\theta>\lambda_u$ (small $Pr$) correspond to different scaling regimes,
and therefore we assign the subscripts $u$ and $\ell$ to regimes I, II, III and IV, which indicate the $u$pper-$Pr$ and $\ell$ower-$Pr$ cases, respectively.
Equating $\epsilon_u$ and $\epsilon_\theta$ to their estimated either bulk or BL contributions and 
employing the classical Prandtl
%%% -Blasius-Pohlhausen  
%%% please here only Prandtl, as it is only about which terms to take into BL equation, not how exactly the ratio goes.
scaling relations for the BL thicknesses  $\lambda_\theta$ and $\lambda_u$ \citep{sch79}, one 
in principle obtains eight theoretically possible scaling regimes.
Regimes II$_u$ and III$_\ell$ are rather small, because, e.g., in II$_u$, it is expected that $\lambda_\theta\geq\lambda_u$ due to the BL-dominance in $\epsilon_\theta$, but on the other hand, $\lambda_\theta\leq\lambda_u$ should hold due to the large $Pr$.
By similar arguments, regime III$_\ell$ is also small.

The mean thicknesses of the BLs are estimated as follows: $\lambda_u\sim L/\sqrt{\text{Re}}$, as in
 RBC \citep{gro00,gro01,Shi2015,Ching2019}, and
$\lambda_\theta\equiv 2/(\lambda_{\theta,\,\text{top}}^{-1}+\lambda_{\theta,\,\text{bottom}}^{-1})$.
Approximating $\left< \partial_z \theta \right>$ at $z=0$ and $z=H$ with the ratio of $\Delta$ and the top and bottom thermal BL thicknesses, $\lambda_{\theta,\,\text{top}}$ and $\lambda_{\theta,\,\text{bottom}}$, from Eqs.~(\ref{tD}), (\ref{eq-n}) and (\ref{eq-n2}) we obtain $\lambda_{\theta}\sim L\tilde\Delta$.

The value of $\epsilon_{u,\text{bulk}}$ is estimated as 
\begin{eqnarray*}
\epsilon_{u,\text{bulk}}\sim 
U\frac{U^2}{L}\frac{L-\lambda_u}{L}\approx\frac{U^3}{L}=\frac{\nu^3}{L^4}Re^3,
\end{eqnarray*}
which is relevant in the $\epsilon_u$-bulk  dominating regimes II$_\ell$, IV$_\ell$ and IV$_u$,
while the value of $\epsilon_{\theta,\text{bulk}}$ is  estimated as 
\begin{eqnarray*}
\epsilon_{\theta,\text{bulk}}\sim 
U\frac{\Delta^2}{L}\frac{L-\lambda_\theta}{L}\approx\frac{U\Delta^2}{L}=\frac{\kappa\Delta^2}{L^2}{Pr\,Re},
\label{above}
\end{eqnarray*}
which is relevant in the $\epsilon_\theta$-bulk  dominating regime IV$_\ell$.
For large $Pr$ (regimes III$_u$ and IV$_u$), the thermal BL is embedded into the kinetic one and therefore in eq. (\ref{above}), the magnitude
of the velocity of the flow, which carries the temperature in the bulk, is reduced from $U$ to $(\lambda_\theta/\lambda_u)U$, leading to
\begin{eqnarray}
\label{51}
\epsilon_{\theta,\text{bulk}}\sim 
\frac{\lambda_\theta}{\lambda_u}
\frac{U\Delta^2}{L}\frac{L-\lambda_\theta}{L}\approx
\frac{\kappa\Delta^2}{L^2}Pr\,Re^{3/2}\,\tilde\Delta.
\end{eqnarray}

The kinetic dissipation rate in the BL is $\sim \nu(U/\lambda_u)^2$. Hence,
\begin{eqnarray}
\label{52}
\epsilon_{u,\text{BL}}\sim 
\nu\frac{U^2}{\lambda_u^2}\frac{\lambda_u}{L}=\frac{\nu^3}{L^4}{Re}^{5/2},
\end{eqnarray}
which is relevant in the $\epsilon_u$-BL  dominating regimes I$_\ell$, I$_u$ and III$_u$.
As in \cite{gro00,gro01}, the factor $\lambda_u/L$ accounts for the volume fraction of the kinetic BL. 
With increasing $Pr$, $\lambda_u$ saturates to $\sim L$, so this factor becomes one (just as argued in \cite{gro01}), 
which yields
\begin{eqnarray}
\label{forstar}
\epsilon_{u,\text{BL}}\sim 
\nu\frac{U^2}{\lambda_u^2}=\frac{\nu^3}{L^4}{Re}^{2}.
\end{eqnarray}
For small $Ra$ or very large $Pr$, this leads to special regimes I$_\infty^<$ and III$_\infty$ on top of, respectively, I$_u$ and III$_u$.
In III$_\infty$, also $\epsilon_{\theta,\text{bulk}}$ scales differently to (\ref{51}), namely as
\begin{eqnarray*}
\epsilon_{\theta,\text{bulk}}\sim 
\frac{\lambda_\theta}{L}
\frac{U\Delta^2}{L}\frac{L-\lambda_\theta}{L}\approx
\frac{\kappa\Delta^2}{L^2}Pr\,Re\,\tilde\Delta.
\end{eqnarray*}

The thermal dissipation rate in the BL scales as $\sim\kappa(\Delta/\lambda_\theta)^2$, which is relevant in the $\epsilon_\theta$-BL  dominating regimes I$_\ell$, I$_u$, I$_\infty^<$ and II$_\ell$. This
(again with the volume fraction factor) 
 leads to 
\be
\epsilon_{\theta,\text{BL}}\sim 
\kappa\frac{\Delta^2}{\lambda_\theta^2}\frac{\lambda_\theta}{L}=
\kappa\frac{\Delta^2}{L^2}\frac{\lambda_u}{\lambda_\theta}{Re}^{1/2}.
\label{epsthetabl} 
\ee
In the limiting regimes I$_\ell$, II$_\ell$ and I$_\infty^<$, it 
holds $\lambda_u/\lambda_\theta\sim {Pr}^{1/2}$ \citep{sch00, shi2017}, while
in regime I$_u$ it holds $\lambda_u/\lambda_\theta\sim {Pr}^{1/3}$, all just as in the classical Prandtl--Blasius--Pohlhausen theory
\citep{sch79}. 
%%% just as -- yes

Equating $\epsilon_u$ and $\epsilon_\theta$ to their estimated bulk or BL contributions,
we obtain the scalings of $\tilde\Delta$ and $Re$ in IHC, which are summarized in Tab.~\ref{Tab1} and sketched in Fig.~\ref{Fig1}.

As already mentioned above, the very same idea was already applied to horizontal convection \citep{shi16}. 
Interestingly  enough, even a formal analogy between IHC and HC exists, out of which we could have already derived
the scaling relations of Table~\ref{Tab1} and Fig.~\ref{Fig1}. 
The reason for this formal analogy is that 
the relations  obtained for $\epsilon_\theta$ and $\epsilon_u$ (see Eqs.~(5) and (6) of \cite{shi16}) 
 formally resemble the corresponding  relations (\ref{ex-th}) and (\ref{ex-u}) here. 
 For the first equation this becomes  particular obvious  when writing 
 $\epsilon_\theta = {L^2 \over \kappa} \Omega^2 \tilde \Delta = {\kappa  \over L^2} \Delta^2 \tilde \Delta^{-1}  $ 
 and for the second when realizing that $( {1\over 2} - \tilde Q_0 )$ is only a dimensionless factor between 0 and 1/2. 
 Then one sees immediately that the role of the control parameter $Ra$ in HC is taken by that of the control parameter
 $R$ in IHC and the role of the response parameter $Nu $ in HC is taken by that of the (inverse) response parameter $\tilde \Delta^{-1}$ in IHC. 
 All derived scaling relations in the different limiting regimes of HC can directly be taken over.  
 The corresponding values for IHC give the same results as obtained above and have already been 
 shown in Table~\ref{Tab1} and Fig.~\ref{Fig1}. 

\begin{figure}
\unitlength1truecm
\includegraphics[width=0.7\textwidth]{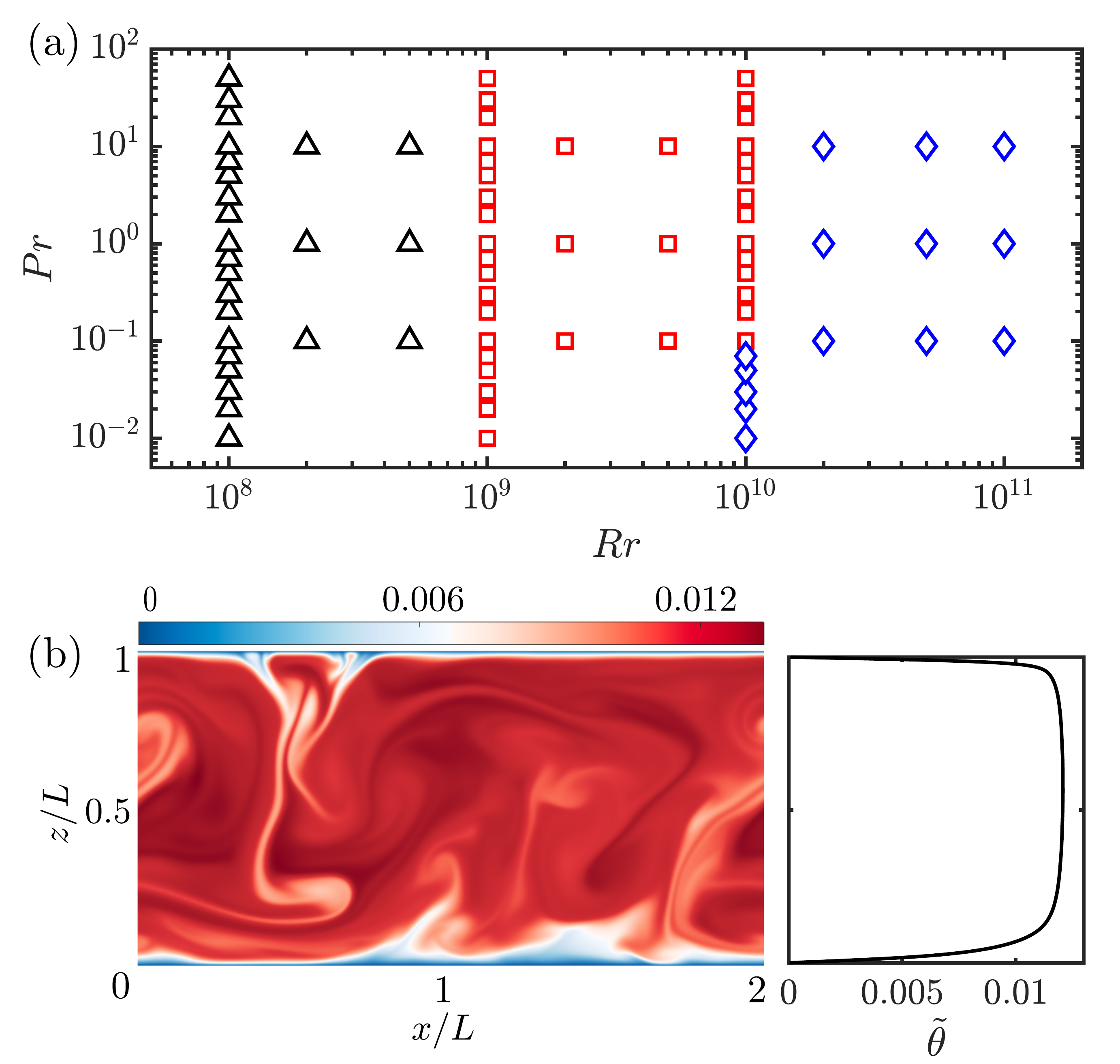}
\caption{
(a) $Rr$ vs. $Pr$ parameter space of the simulated cases. 
Symbols denote the different grid resolutions used in DNS: $512\times256$ ({\color{black}{$\triangle$}}), $1024\times512$ ({\color{red}{$\Box$}}), $2048\times1024$ ({\color{blue}{$\Diamond$}}).
(b) Instantaneous temperature field (color coded) and mean temperature profile for $Rr=10^{10}$ and $Pr=1$. 
}
\label{Fig2}
\end{figure}

\begin{figure}
\unitlength1truecm
\includegraphics[width=1.0\textwidth]{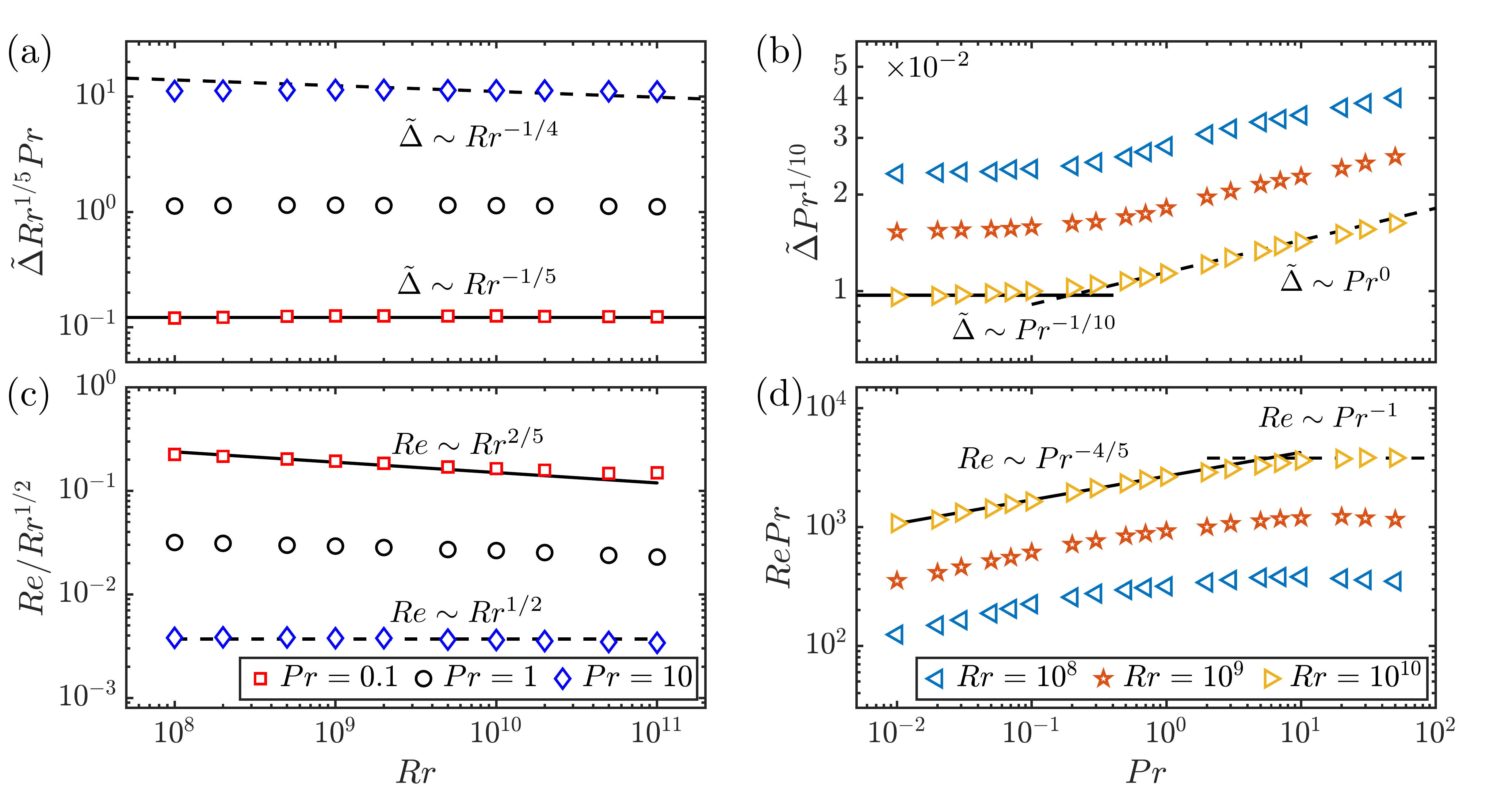}
\caption{Response parameters $\tilde \Delta $ (the dimensionless mean temperature of the bulk)
and $Re$ as function of the control parameters $Rr$ and $Pr$:  
(a) Compensated $\tilde \Delta$ as function of $Rr$ for fixed $Pr = 10^{-1}$, $1$, $10$.  
(b) Compensated $\tilde \Delta$ as function of $Pr$ for fixed $Rr= 10^8$, $10^9$, $10^{10}$. 
(c) Compensated $Re$ as function of $Rr$ for fixed $Pr = 10^{-1}$, $1$, $10$.  
(d) $Re Pr$ as function of $Pr$ for fixed $Rr= 10^8$, $10^9$, $10^{10}$. 
The straight lines with the corresponding scaling laws are added as guide to the eye. 
%Compensated three-dimensional visualization of (e) $\tilde{\Delta}(Rr,Pr)$ and (f) $Re(Rr,Pr)$ 
}
\label{Fig3}
\end{figure}

\begin{figure}
\includegraphics[width=1\textwidth]{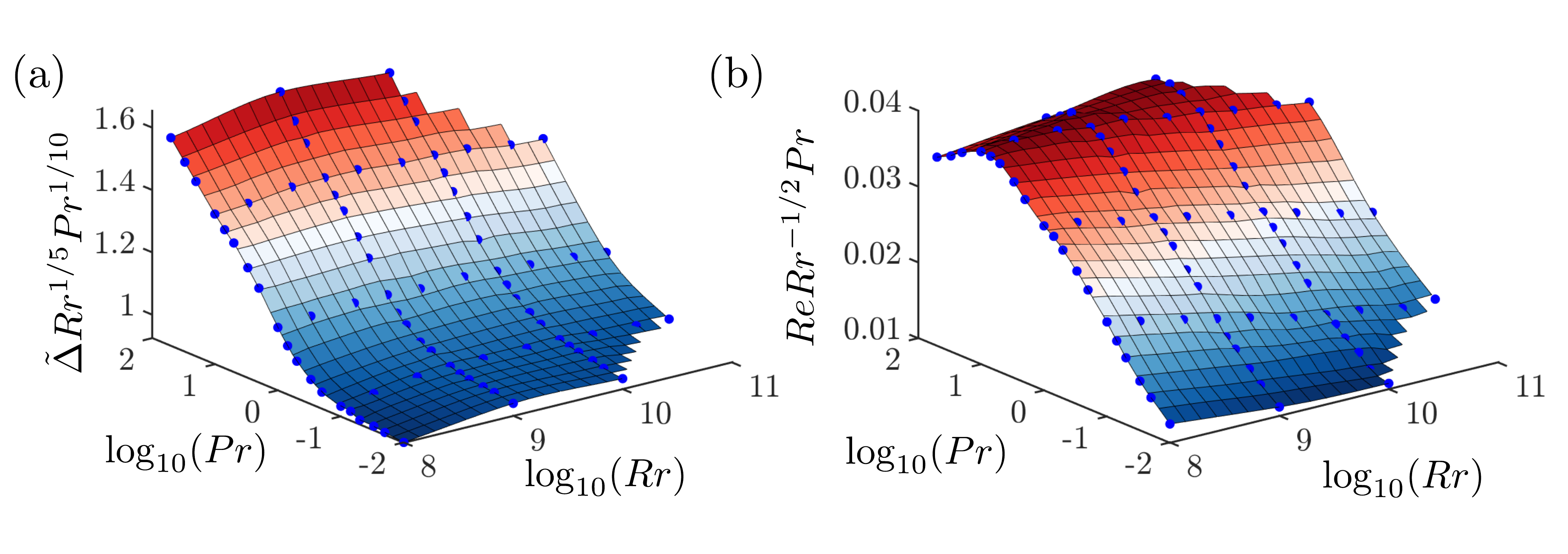}
\caption{Compensated three-dimensional visualization of (a) $\tilde{\Delta}(Rr,Pr)$ and (b) $Re(Rr,Pr)$.}
%3D visualizations of the same  response parameter as in Fig.~\ref{Fig3}.
\label{Fig4}
\end{figure}

To check these predictions of the GL theory generalized to IHC, we have performed 2D DNS according to Eqs.~(\ref{eq1}) and (\ref{eq2}) with the corresponding BCs. We chose an aspect ratio of $\Gamma = 2$ for the laterally periodic box. 
The numerics have been validated by making sure that the exact relations (\ref{ex-u}) are fulfilled. Simulations were performed using the second-order staggered finite difference code AFiD \citep{verzicco1996,van2015}. This code has already been extensively used to study RBC, see, e.g. 
\citep{wang2020,wang2020zonal}.

The parameter combinations ($Rr$, $Pr$) for which we performed simulations are shown in the parameter space of Fig.~\ref{Fig2}a. 
A typical snapshot of the temperature field together with the mean temperature profile for one parameter combination are displayed in Fig.~\ref{Fig2}b. 
One can see the stably-stratified layer near the bottom plate. 
The interaction of the upper convection zone and the lower stably-stratified  region leads to the so-called penetrative convection \citep{veronis1963,wang2019}.
 The mean temperature profile, which, as expected and  typical for IHC, displays top-down asymmetry. 

The results for the response parameters $\tilde \Delta $ and $Re$ as functions of the control parameters $Rr$ and $Pr$ 
are shown in Figs.~\ref{Fig3} and \ref{Fig4}. 
As can be seen, in general, there are  no pure scaling laws over the simulated range, but smooth crossovers 
from one regime to the other, very similarly as in RBC \citep{ste13}, reflecting the key idea of the unifying theory by \cite{gro00,gro01}. 
We first discuss the dependences for the dimensionless mean temperature $\tilde \Delta$, see Figs.~\ref{Fig3}a,~b. 
As a function of $Pr$ (Fig.~\ref{Fig3}b), for all $Rr$ the transition from $\tilde \Delta \sim Pr^{-1/10}$
of regime $I_l$ to the $Pr$-independence of regime $I_\infty^{<}$ can clearly be seen. The more turbulent regimes
$IV_{u,\ell}$ are not yet realized, as the driving is not strong enough. 
This is also reflected in the $Rr$ dependence $\tilde {\Delta} \sim Rr^{-1/5}$ reflecting that of regimes $I_{u,\ell}$. 
No indication to a stronger dependence as typical for the more turbulent regimes $IV_{u,\ell}$ can yet be seen. 
This is also seen in the dependences of the Reynolds number (Fig.~\ref{Fig3}~c,~d): 
For small $Pr \le 1$, it goes like $Re \sim Rr^{2/5}$ as in regimes $I_{u,\ell}$. 
For large $Pr=10$ the results are consistent with $Re\sim Rr^{1/2}$ as in regime $I_\infty^{<}$. 
This scaling should go hand in hand with the scaling $\tilde {\Delta} \sim Rr^{-1/4}$ for the dimensionless mean temperature, but as seen from Fig.~\ref{Fig3}a, those data are presumably better described by $\tilde {\Delta} \sim Rr^{-1/5}$. 
Finally, on the $Pr$-dependence of $Re$: 
As seen from Fig.~\ref{Fig3}d, for all $Rr$ the data show a transition from the $Re\sim Pr^{-4/5}$ scaling of regimes $I_{u,\ell}$ to the $Re\sim Pr^{-1}$ scaling of regime $ I_\infty^<$, consistent with  the corresponding transition for $\tilde \Delta $
in Fig.~\ref{Fig3}b. 

\begin{figure}
\unitlength1truecm
\includegraphics[width=1\textwidth]{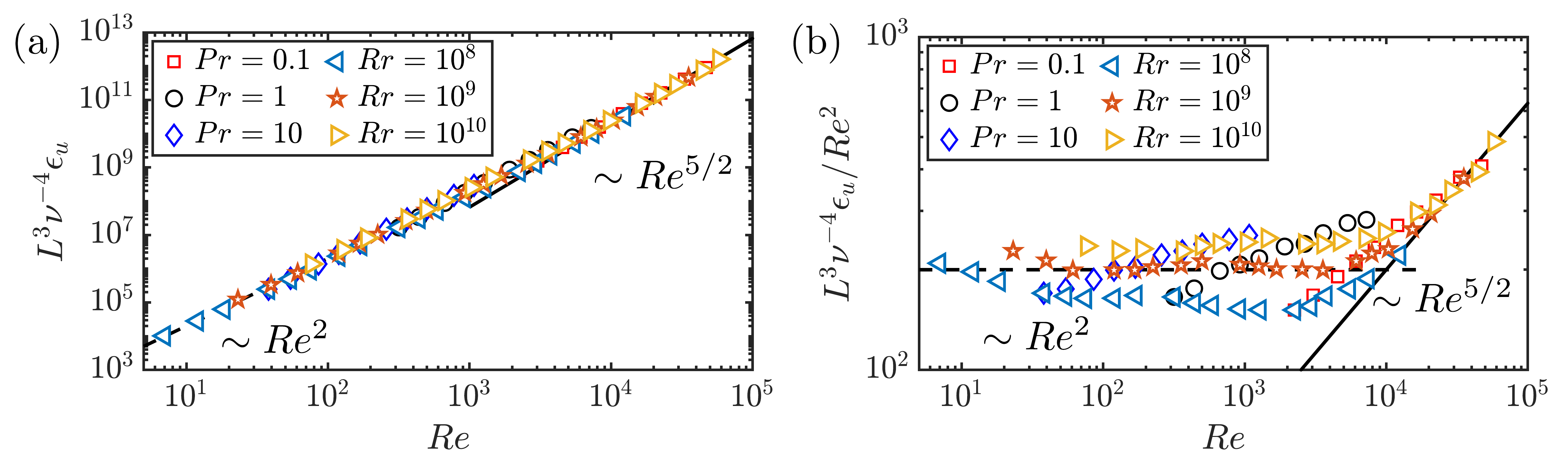}
\caption{The non-dimensionalized (a) absolute kinetic energy dissipation rate 
$L^3\nu^{-4}\epsilon_u$ and (b) compensated kinetic energy dissipation rate $L^3\nu^{-4}\epsilon_u/Re^2$ as function of $Re$.
Note that $L^3\nu^{-4}\epsilon_u \sim Rr Pr^{-2}$, Eq.~\ref{ex-u}, holds exactly throughout 
and has numerically been checked for consistency (not shown). Also note that 
the non-dimensionalized thermal energy dissipation rate (middle column of Table~\ref{Tab1}) is nothing else
but 
%the inverse of 
$\sim\tilde \Delta^{-1}$ and has thus already been shown in Figs.~\ref{Fig2}~a,~b. 
%\red{Qi, please plot as function of Re}. 
}
\label{Fig5}
\end{figure}

\begin{figure}
\unitlength1truecm
\includegraphics[width=1\textwidth]{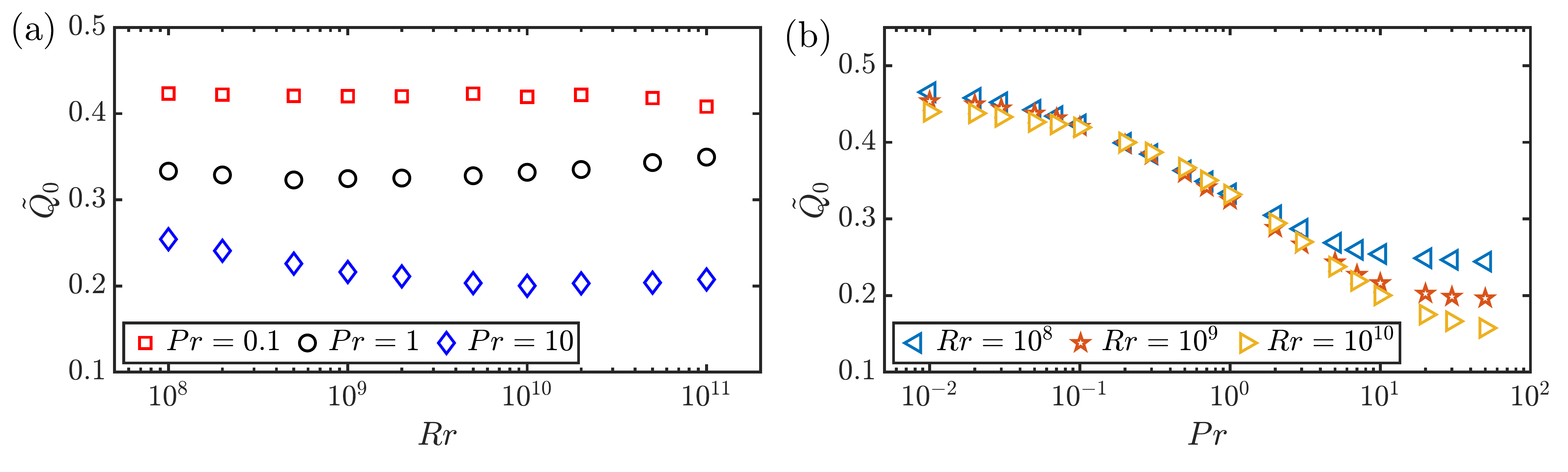}
\caption{
Magnitude of the dimensionless heat flux through the bottom plate, $\tilde Q_0$, as function of (a) $Rr$ for the different $Pr$ and (b) $Pr$ for the different $Rr$ (also shown in legends). 
}
\label{Fig6}
\end{figure}

All these results are consistent with our unifying theory, which however goes much beyond the simulated 
parameter range into the regimes in which the kinetic and thermal energy dissipation rates are dominated by
the turbulent bulk contributions. These regimes are inaccessible with our present numerical simulations, even in 2D. 

As an additional check of our unifying theory we also plot the kinetic energy dissipation rate as function of $Re$, 
see Fig.~\ref{Fig5}. 
Indeed, we find $\eps_u/ (L^3 \nu^{-4}) \sim Re^{5/2}$ and $\sim Re^2$ as characteristic for the kinetic BL dominated regimes $I_{u,\ell}$ and $I_\infty^{<}$, consistent with what we have seen in Fig.~\ref{Fig2}. 

Another (less important) response parameter of IHC is the magnitude of the dimensionless heat flux $\tilde Q_0$ through the bottom plate.
The numerical results for $\tilde Q_0$ are shown in Fig.~\ref{Fig4}. 
One sees from Fig.~4a that $\tilde Q_0$ only weakly depends on $Rr$ in the present parameter range; this behaviour has also been found before in \cite{goluskin2016penetrative}. 
Fig.~4b illustrates that much less heat is transported outwards from the bottom plate with increasing $Pr$. 
The small $\tilde{Q}_0$ for large $Pr$ is due to the less efficient shear-driven mixing of the fluid near the bottom plate.

In conclusion, in the spirit of the prior unifying theories for 
RBC \citep{gro00,gro01} and for HC \citep{shi16}, in this paper we have developed a unifying
 theory of IHC  for the scaling of the mean temperature and the Reynolds
number as functions of the control parameters $Rr$ and $Pr$.  
The main result is visualized in Fig.~\ref{Fig1}. 
We have shown that the 2D DNS results are consistent with this theory, though the  numerically
accessible regimes are still dominated by the boundary layers, and not all predictions of the theory can already
be tested at this point. Also 3D DNS are still outstanding. We have furthermore pointed towards the formal analogy
between IHC and HC and it will be illuminating to explore this analogy also numerically.

%\red{Connect back to introduction and put results in broader context. Also give outlook to other cases.}

%\begin{acknowledgments}
{\it Acknowledgements:}  
R.~Verzicco, D.~Goluskin and K.~L.~Chong are gratefully acknowledged for discussions and support. 
We  also acknowledge the Twente Max-Planck Center,
the Deutsche Forschungsgemeinschaft (Priority Programme SPP 1881 "Turbulent Superstructures").
%PRACE for awarding us access to MareNostrum based in Spain at the Barcelona Supercomputing Centre (BSC) under PRACE project number 2017174146. 
The simulations were carried out on the national e-infrastructure of SURFsara, a subsidiary of SURF cooperation, the collaborative ICT organization for Dutch education and research. Q.W. acknowledges financial support from China Scholarship Council (CSC) and Natural Science Foundation of China under grant no. 11621202.
%\end{acknowledgments}

D.L. and O.S. contributed equally to this work.

\bibliographystyle{prsty_withtitle}
\bibliography{literatur}

\end{document}